%% file: IJUC_050618_Cococcioni.tex
\documentclass[a4paper,11pt]{ijuc} 

\usepackage[USenglish]{babel} 
\usepackage[T1]{fontenc}
\usepackage[ansinew]{inputenc}

\usepackage[title]{appendix}

\usepackage{lmodern} 

\usepackage{comment}

\usepackage{graphicx} 
\usepackage{placeins}

\usepackage{amssymb}
\usepackage{pifont}

\usepackage{stmaryrd}

\makeatletter
\renewcommand{\@biblabel}[1]{#1.}
\makeatother

\def\G1{\hbox{$\displaystyle{\mbox{\ding{172}}}$}}

\newcommand{\inn}{\! \in \!}

\usepackage{soul}

\usepackage{color}

\usepackage{amsfonts, amsmath,amsthm}

\newtheorem{numbered_theorem}{Theorem}

\DeclareMathOperator*{\argmax}{\arg\!\max}

\begin{document}



\title{Numerical Asymptotic Results \\
in Game Theory Using \\
Sergeyev's Infinity Computing}

\author{Lorenzo Fiaschi \inst{1}\email{lorenzo.fiaschi@gmail.com}
\and Marco Cococcioni\inst{1}\email{marco.cococcioni@unipi.it} 
}

\institute{Department of Information Engineering,
University of Pisa\\Largo Lucio Lazzarino 1 -- 56122 Pisa, Italy
}

\maketitle

\begin{abstract}
Prisoner's Dilemma (PD) is a widely studied game that plays an important role in Game Theory.
This paper aims at extending PD Tournaments to the case of infinite, finite or infinitesimal payoffs
using Sergeyev's Infinity Computing (IC).
By exploiting IC, we are able to show
the limits of the classical approach to PD Tournaments analysis of the
classical theory, extending both the sets of the feasible and numerically computable tournaments.
In particular we provide a numerical computation of
the exact outcome of a simple PD Tournament
where one player meets every other an infinite number of times,
for both its deterministic and stochastic formulations.


\end{abstract}



\section{Introduction}
\label{intro}
Nowadays Game Theory (GT) is widely used in very different areas,
such as biology, engineering, social sciences, 
and so on. 
An important role in matrix GT is played 
by the well-known Prisoner's Dilemma (PD) 
and its variants.
This kind of games continue to attract the interest of researchers,
especially under challenging settings like infinitely repeated games \cite{Burger_12}, 
games with an infinite amount of players \cite{Sabak_05} 
or games with infinitesimal probabilities \cite{Blume_91}.

Despite these remarkable attempts to extend the 
GT to the case of infinite 
or infinitesimal quantities, most of the times 
conditions are imposed to reduce the original problem to one involving finite quantities.
This approach is typical to limit the value of the payoffs to make them
finite (and thus comparable) quantities \cite{Reny_99, Chambers_09, Rowat_07}, allowing
one to resort to the familiar
machinery 
of classical GT with finite quantities.
Real applications, on the contrary, sometimes involve 
payoffs which are incomparably larger
than others (e.g., losing own life \emph{vs} losing one dollar).

Though in the past we had no
numerical tools suitable to
deal with
infinite, finite and infinitesimal quantities, this is 
no longer the case since the introduction
of Sergeyev's Infinity Computing (IC), happened in 2003 \cite{Sergeyev}.

Thanks to this novel approach, many open problems have been tackled
using it by performing \textit{numerical} computations 
with infinite and infinitesimal 
quantities in a handy way (for detailed introductory surveys see 
\cite{EMSS_17,informatica,Lagrange,NewZealand,UMI}, while 
\cite{Sergeyev} contains an introduction to the topic written in a popular way).
This computational methodology has already been successfully applied in optimization
and numerical differentiation \cite{Cococcioni_16,Cococcioni_18, Mukhametzhanov_18,Num_dif,Sergeyev_18,Zilinskas} and in a number of other
theoretical and computational research areas such as cellular
automata \cite{DAlotto_2}, Euclidean and hyperbolic 
geometry \cite{Margenstern_3}, 
fractals \cite{Caldarola_18, Caldarola_18_bis, Biology,Koch,DeBartolo}, Bertrand's Paradox and mathematical determination \cite{Rizza_17}, 
Turing machines \cite{Sergeyev_Garro_2, Sergeyev_Garro_1},
and numerical solution of  ordinary differential equations \cite{ODE_3, Sergeyev_16}.

This methodology uses a numeral system working with an infinite
number called \emph{Grossone}, expressed by the numeral \G1, and
defined as the number of elements of the set of natural numbers
(the consistency, and more precisely the relative consistency 
of the methodology, has been studied in \cite{Lolli_2}).
This numeral system allows to denote a variety
of numbers involving different infinite and infinitesimal parts and
to execute operations with all of them in a \emph{single} framework.
Such numeral system and the associated operations can be implemented 
in hardware for a new generation of computers, called 
by Sergeyev himself \emph{Infinity Computers}.


We have exploited Sergeyev's IC for studying PD 
Tournaments.
In particular we have been able to 
overcome the
issues that occur using the classical 
limit theory for studying a player 
that meets the others an infinite number 
of times. 






In Section 2 we briefly describe 
the classical formulations 
of PD Tournaments.
Then in Section 3 we introduce the IC
and the associated Grossone Methodology (GM). 
In Section 4 and 5 we extend
PD Tournaments in both deterministic 
and stochastic settings, 
using the GM. 
Section 6 presents the numerical results of the GM applied to stochastic PD Tournaments, 
while Section 7 is devoted to conclusions.

\section{Classical Formulation of PD Tournaments}
In the generalized form of the PD two players have 
to independently decide whether to
cooperate ($C$) or to defect ($D$). The first player 
will get one of the following payoffs
$T_1, R_1, P_1,$ or $S_1$, depending on his and other player's choice.
Similarly, the second player will get $T_2, R_2, P_2,$ or $S_2$.
In this paper we will assume the games are symmetric, i.e., $T_1=T_2=T$, $R_1=R_2=R$, etc.
In addition a game is called a PD game if the payoffs follow the relation $T>R>P>S$ (\emph{fundamental law}) and the payoff obtained by a player is $T$, $R$, $P$ or $S$ when the outcome is $DC, CC, DD,$ or $CD$, respectively. The latter condition reads as follows: a player gets the reward $T$ when he has chosen to defect and the other has chosen to cooperate, while he gets $R$ when they have both cooperated, and so on. Further information can be found in \cite{Stanford}.

\subsection{PD Tournament}
The basic PD game described above can also be repeated many times among multiple players. In the literature, when a PD game is repeated multiple times, it is denoted as \emph{iterated} PD.
A PD Tournament is a particular type of iterated PD, characterized by
a set of players, where each one has his own strategy $\mathcal{S}$.
Every player meets each other a large number of times and at each interaction 
the basic (one-shot) PD game is played, then the reward 
of a player is summed up to his own previous ones.
The player with the greatest total reward is the \emph{tournament winner}, and the associated strategy is the \emph{winning strategy}.
The strategy of each player can be defined as the probability of cooperation at each step of the repeated game, taking into account what the other players have done in the previous games (if they have cooperated or not).

Many different types of tournament exist. In this paper we focus on tournaments 
with memory-one strategies in a non evolutionary context, i.e., tournaments where: 
i) the strategies at a given stage are influenced by the outcome of the previous stage only, and ii) there are no concepts of population (i.e.
each strategy is represented at most once).

In this setting the strategy $\mathcal{S}$ adopted by each player can be described by a tuple of five numbers, $(y, p_1, p_2, p_3, p_4)$, 
where $y$ is the probability to cooperate at the first
round and the four values $p_1,..., p_4$ 
are the probabilities to cooperate if the outcome of the previous round was, respectively, $CC, CD, DC,$ or $DD$
($p_1$ represents the probability for cooperating during next
interaction, provided that both of the players have cooperated in the previous interaction, $p_2$ is the probability of cooperating if the player has cooperated while the other did not, and so on).

Notice that the memory-one assumption in many cases does not constitute a limitation, due to 
the results provided in \cite{Hilbe_13, Press_Dyson_12, Stanford}.

In Section 4 and 5 we will generalize these PD tournaments to the
case of infinite or infinitesimal payoffs and after infinite interactions. Before doing that, let us introduce the mathematical tools for describing infinite and infinitesimal numbers.

\section{Infinity Computing and Grossone \\ Methodology}

As said before, in
\cite{Sergeyev,informatica,Lagrange,UMI,NewZealand} a  computational
methodology able to deal with infinite, finite and infinitesimal
numbers in the same single framework has been realized by means of a new numeral system with infinite
base. The latter is called \emph{Grossone}, indicated by the numeral $\G1$
and defined as the number of elements in the set of natural numbers
$\mathbb{N}$.

The new numeral \G1 has been introduced by describing its properties (following
the same approach that led to the introduction of the zero in the
past to switch from natural to integer numbers). To introduce
\emph{Grossone}, a methodological platform has been populated with three methodological postulates and The Infinite
Unit Axiom is added to the axioms of real numbers \cite{EMSS_17,Sergeyev, Lagrange, NewZealand,UMI}. In particular, the
axiom states that for any given finite integer $n$ the
number $\frac{\small{\G1}}{n}$ is natural and infinitely large, since $\frac{\small{\G1}}{n} < \G1$.
More precisely, $\frac{\small{\G1}}{n}$ denotes the number of elements of any arithmetical progression
of the form $k$, $k + n$, $k + 2n$, $\ldots$, with $1 \leq k \leq n$ and $k, n$ finite \cite{Rizza_17}.
Since the axiom is added to the standard axioms of real
numbers, all standard properties (commutative, associative,
existence of inverse, etc.) also apply to \G1 and
\emph{Grossone}-based numbers.  Instead of the usual symbol
$\infty$, different infinite and/or infinitesimal numbers can be
used thanks to \G1. Indeterminate forms as encountered in the classical analysis (e.g., $\infty-\infty$, $\frac{\infty}{\infty}$)
are no more present and, for
example, the following relations hold for any finite, infinite and infinitesimals number expressible in the
new numeral system:
\vspace{+0.5mm}
\[
 0 \cdot \mbox{\ding{172}} =
\mbox{\ding{172}} \cdot 0 = 0, \hspace{3mm}
\G1^{-1} >   \G1^{-2} > 0, \hspace{3mm}
2\G1-\G1= \G1\]
\[\G1^0 = \G1^{1-1} = \G1^1 \cdot \G1^{-1} = \frac{\small{\G1}}{\small{\G1}} = 1, \hspace{3mm} \G1^{-1} = \G1 \cdot \G1^{-2} = \frac{\small{\G1}}{\small{\G1^2}}.
\]
\vspace{0.5mm}

In particular, \emph{Grossone} works as a base in a numeral system whose generic element
$\tilde{c}$ 
(called \emph{gross}-scalar) can be represented with a notation in the middle between the polynomial and the common 
positional numeral systems one:

\vspace{0.5mm}
\[
\tilde{c}=c_{p_m}\G1^{p_m} + ... + c_{p_1}\G1^{p_1} +
c_{p_0}\G1^{p_0} + c_{p_{-1}}\G1^{p_{-1}} + ... +
c_{p_{-k}}\G1^{p_{-k}} ,
\]
\vspace{0.5mm}

where $m, k \inn \mathbb{N}$, exponents $p_i$ are called
\emph{gross}-powers (they can be numbers of the type of $\tilde{c}$)
with ${p_0} = 0,$ and $ i=m,...,1,0,-1,...,-k$. Then,
 ${c_p}_{_i}\neq 0$ called \emph{gross}-digits are
finite (positive or negative) numbers,   $ i=m,...,1,0,-1,...,-k$.
In this numeral system, finite numbers are represented by
numerals with the highest \emph{gross}-power equal to zero, e.g.,
$-6.2=-6.2\G1^0$. Infinitesimals are represented by numerals having only
negative (finite or infinite) \emph{gross}-powers. The simplest
infinitesimal is $\G1^{-1}$ for which $\G1^{-1} \cdot \G1 = 1$. Moreover
it is worth noting that all infinitesimals are not equal to zero, e.g.,
$\G1^{-1}
 >0$. Finally, a number is infinite if it has at least one positive finite or
infinite gross-power. For instance, the number
$43.6\G1^{4.56\mbox{\tiny{\ding{172}}}}+16.7\G1^{3.6}-3.2\G1^{-2.1}$
is infinite, it consists of two infinite parts and one infinitesimal
part.

The new methodology provides a computational framework that handles
infinite and infinitesimal numbers
and a floating-point arithmetic can be carried out with gross-numbers that have determinate infinite and infinitesimal values.

These latter properties set Sergeyev's methodology apart from non-standard analysis \cite{Robinson}, where non-standard
infinite and infinitesimal numbers are introduced but there is no
way of assigning them specific and concrete values, no way to manage them numerically
by a calculator, no way to instantiate ordering relationships among
them. Indeed  non-standard analysis
is a purely symbolic technique that works with ultrafilters, external and internal sets,
standard and non-standard numbers. 

On the other hand, the IC-based approach does not use
any of these notions, focuses on numerical computations and separates mathematical
objects from tools used to study them,
being sensitive to the fact that the instruments employed both determine the kind of possible interventions on the object of study and constrain the accuracy of the numerical results that can be obtained.

The foregoing remarks, as well as other reasons (e.g., the possibility to establish the maximal discriminable number
in a sequence) show how the \emph{Grossone}-based
methodology differs from other kinds of infinitesimal methods. Further differences are discussed in \cite{G_NSA_18}.


\section{Deterministic PD Tournament using \\ Grossone Methodology}

In Deterministic PD Tournaments (DPDTs) \cite{Axelrod_84, Lindgren_91}
with $m$ players, every strategy $\mathcal{S}_k(y_k, p_{1k}, p_{2k}, p_{3k}, p_{4k})$
has parameters that are either equal to 0 or equal to 1:

\vspace{0.5mm}
\[y_k, \, p_{1k}, \, p_{2k}, \, p_{3k}, \, p_{4k} \in \{0,1\} \quad \forall k = 1,...,m  \, .
\]
\vspace{0.5mm}

Let us to consider a 
classical DPDT 
with three players $\wp_1$, $\wp_2$ and $\wp_3$, 
identified by their own strategies:  $Du,$
$TRIGGER$ and $TFT$ defined as follows:
\[
\vspace{0.5mm}
\begin{matrix}
Du = \mathcal{S}(0, 0, 0, 0, 0) \\
{}\\
TRIGGER = \mathcal{S}(1, 1, 0, 0, 0) \\
{}\\
TFT = \mathcal{S}(1, 1,  0, 1, 0).
\end{matrix}
\]
\vspace{0.5mm}
The three strategies work as follows.
\vspace{0.5mm}
\begin{itemize}
\item The $Du$ strategy consists in a constant defection.
\item The $TRIGGER$ strategy (also known as GRIM
TRIGGER or simply GRIM \cite{Gibbbons_92,Stanford,Szabo_07}) 
starts cooperating, then the cooperative behavior stands as long as both players cooperate. At the first opponent's uncooperative choice it will never cooperate anymore.
\item Finally the $TFT$ strategy (\emph{tit-for-tat}, the strategy proposed by the professor Anatol
Rapoport for the Axelrod's PD Tournaments \cite{Axelrod_84, Axelrod_81}) starts cooperating and then emulates the opponent's previous behavior.
\end{itemize}
\vspace{0.5mm}

Starting the tournament, after each strategy has been played against each other $n$ times,
we obtain the results given in Table~\ref{tab:tab1}. There, $E_{\wp_i, \wp_j}^{\;n}$ is the short form of $E_{\wp_i, \wp_j}(n)$ and reads as the Expectation of $\wp_i$ when plays against $\wp_j$ $n$ times; $E_{\wp_i}^{\;n}$ stands for $E_{\wp_i}(n)$ and reads as the total expectation of the player $\wp_i$ after having played against each other
player $n$ times. A complete explanation of these result is provided in the Appendix A.

\vspace{0.5mm}
\begin{table}[bht]
\begin{center}
\footnotesize{        
		\begin{tabular}{p{3cm} | p{3.3cm} | p{3.3cm}}
		\hline        
		\hspace{1cm} $\wp_1$ & \hspace{1cm} $\wp_2$ & \hspace{1cm}  $\wp_3$\\
		\hline
        & & \\
        $E_{\wp_1,\wp_2}^{\;n}$ = $T+ (n-1)P$  &  $E_{\wp_2,\wp_1}^{\;n}$ = $S+ (n-1)P$ &  $E_{\wp_3,\wp_1}^{\;n}$ = $S + (n-1)P$\\
        && \\
		$E_{\wp_1,\wp_3}^{\;n}$ = $T+ (n-1)P$   &  $E_{\wp_2,\wp_3}^{\;n}$ = $nR$ & $E_{\wp_3,\wp_2}^{\;n}$ = $nR$ \\
        & & \\
        \hline
        & & \\
		$E_{\wp_1}^{\;n}$ = $2T + 2(n-1)P$         &  $E_{\wp_2}^{\;n}$ = $nR+(n-1)P+S$ &
		$E_{\wp_3}^{\;n}$ = $nR+(n-1)P+S$\\
        & & \\
		\hline
		\end{tabular}
}
\end{center}
\vspace{-2mm}
\caption{Expectations (E) of $\wp_1$, $\wp_2$ and $\wp_3$ after $n$ interactions with each other player (the last row is the sum of the two terms on the previous one)}
\label{tab:tab1}
\end{table}
\FloatBarrier
For next discussions, it suffices to focus on players $\wp_1$ and $\wp_2$ and on their performance-wise relation, described by the sign of the function:

\vspace{0.5mm}
\[\Delta(\wp_1, \wp_2, n) = E_{\wp_1}(n)-E_{\wp_2}(n) = 2T + (n-1)P -n R - S. 
\]
\vspace{0.5mm}

Let us study the relation ``$\wp_1$ does better than $\wp_2$ when 
$n\rightarrow\infty$'', having already fixed some finite values for $T$, $P$ and $S$.
In what follows we will compare the result provided by an application of classical analysis those obtained by means of GM.

First of all we state the problem more formally:

\vspace{0.5mm}
\[
\Delta(\wp_1, \wp_2, n) > 0 \hspace{5mm} \textnormal{iff}  \hspace{5mm}  2T + (n-1)P - n R - S > 0 \vspace{1mm}
\]
\vspace{0.5mm}

from which we deduce:

\vspace{0.5mm}
\begin{equation}
\Delta(\wp_1, \wp_2, n) > 0  \hspace{5mm} \textnormal{iff}  \hspace{5mm} R < \frac{2T+(n-1)P-S}{n}
\label{eq:Gamma_Det}
\end{equation}
\vspace{0.5mm}    

If we now study the asymptotic behaviour of this tournament when $n \to +\infty$, using the classical limit theory, we obtain:

\vspace{0.5mm}
\[
\lim_{n \to \infty} \Delta(\wp_1, \wp_2, n) > 0 \hspace{5mm} \textnormal{iff} \hspace{5mm} 
 R < \lim_{n \to \infty} \frac{2T+(n-1)P-S}{n}
\]
\vspace{0.5mm}

from which we obtain:

\vspace{0.5mm}
\begin{equation}
\lim_{n \to \infty}\Delta(\wp_1, \wp_2, n) > 0  \hspace{5mm} \textnormal{iff} \hspace{5mm} R < P.
\end{equation}
\vspace{0.5mm}

Unfortunately, the latter result ($R<P$) breaks the fundamental law of
PD Tournaments: $T > R > P > S$.

Thus, using the classical analysis, we are inclined to conclude that there is no way to build a PD Tournament where $T > R > P > S$ and where ``$\wp_1$ does better than $\wp_2$'' with certainty, when $n \to \infty$ (i.e., asymptotically).

By contrast, using Sergeyev's IC, we are able to build
a PD Tournament that does not break the fundamental law.

Indeed, by simply putting $n=\G1$ in \eqref{eq:Gamma_Det}, we obtain:

\vspace{0.5mm}
\begin{equation}
R < \frac{2T+(\G1-1)P-S}{\G1}
\vspace{2mm}
\end{equation}
\vspace{0.5mm}

which is interesting because $\frac{2T+\left(\small{\G1}-1\right)P-S}{\small{\G1}}$ is greater than $P$, as shown below:

\vspace{0.5mm}
\begin{equation}
\frac{2T+(\G1-1)P-S}{\G1} = P \, + \, (2T-P-S)\G1^{-1} > P
\label{eq:GrossTournamentCondition}
\end{equation}
\vspace{0.5mm}

The inequality in ~\eqref{eq:GrossTournamentCondition} holds because the fundamental law assures that $(2T-P-S) > 0$ (and thus $(2T-P-S)\G1^{-1}$ is a positive quantity, although infinitesimal).

This means that if we choose $R$ within the open interval

\vspace{0.5mm}
\begin{equation}
R \in \left( P \; , \; P \; + \, (2T-P-S)\G1^{-1} \right)
\label{eq:interval}
\end{equation}
\vspace{0.5mm}

we are guaranteed that ``$\wp_1$ does better than $\wp_2$'', asymptotically, i.e., after $\G1$ iterations.

Such an open interval is interesting because its width is infinitesimal, meaning that $R > P$ but $R$ is infinitely close to $P$.

This highlights that the problem is mainly with the value that we can assign to $T, R, P$ and $S$, more than with the use of classical limit theory itself.

Sergeyev's IC allows us to deal with PD Tournaments where payoffs can be infinite, finite, or infinitesimal, or infinitely close to each other.

A similar result could be obtained using Non-Standard Analysis (NSA) as well but, since NSA is a symbolic tool, it cannot be used  to perform \emph{numerical} computations in environments like Matlab, R, Julia, etc, not being able to assign to the non-standard infinities and infinitesimals a concrete number. The possibility to build PD Tournaments and to compute the expectations numerically in Matlab is important when computations become more complex, as in Stochastic PD Tournaments introduced in section 5.

To start appreciating the numerical added-value of IC in this context,
let us pick a numerical value for $R$ within the allowed interval provided in Equation~\eqref{eq:interval}. For instance, we can choose $R$ as its midpoint:

\vspace{0.5mm}
\[
R = P \, + \, \frac{(2T-P-S)\G1^{-1}}{2}
\]
\vspace{0.5mm}

Doing so, we satisfy the following requirements simultaneously:

\vspace{0.5mm}
\begin{itemize}
\item $R > P$ (fundamental law), and 
\item $R < P + (2T-P-S)\G1^{-1}$ (tournament existence condition).
\end{itemize}
\vspace{0.5mm}

Even more importantly, we are able to compute the exact expectations in a \emph{numerical} way
($\wp_1$ always beats $\wp_2$, since $E_{\wp_1}(n) > E_{\wp_2}(n)$ \; $\forall \; n \in \mathbb{N})$:

\vspace{0.5mm}
\[
  E_{\wp_1}(\G1) = 2P\G1+2(T-P), \;\;\;\;\;\; E_{\wp_2}(\G1)= \G1(R+P) - P + S,
\]
\vspace{0.5mm}

and $E_{\wp_1}(\G1) > E_{\wp_2}(\G1)$ since $2T > P+S$ (due again to the fundamental law).

Thus we have built a tournament where $R$ is a function of $T$, $P$ and $S$. If we now assign values to $T$, $P$ and $S$, we can build and solve a specific tournament as in the following instance:

\vspace{0.5mm}
\[
\begin{cases}
   T = 10  \\
   R = 4+8.5\G1^{-1}  \\
   P = 4  \\
   S = -1  \\
\end{cases}
\vspace{3mm}
\]

The players final ranking is therefore:


\begin{table}[h]
\begin{center}
\begin{tabular}{lllll}
 1\textsuperscript{st} & $\wp_1$  & & $E_{\wp_1}(\G1) = 8\G1+12 $ & \\
 2\textsuperscript{nd} & $\wp_2, \; \wp_3$ & & $E_{\wp_2}(\G1) = E_{\wp_3}(\G1) = 8\G1 + 3.5$ &  \textnormal{(ex-aequo)} \\
\end{tabular}
\end{center}
\end{table}
\FloatBarrier

As a final remark before concluding this section, let us observe how the PD Tournament built above assures that
``$\wp_1$ does better than $\wp_2$'' not only asymptotically, but also when the number of interactions assumes a finite value, i.e., \emph{always}. This means that we have been able to build a PD Tournament where ``$\wp_1$ does better than $\wp_2$'', \emph{irrespective of the number of interactions}.

\section{Stochastic PD Tournaments  using\\
Grossone Methodology}

Stochastic PD Tournaments (SPDTs) \cite{Nowak_90, Nowak_90b} are a generalization of the previously introduced DPDTs. 
In particular they are tournaments where one or more 
players use a non-deterministic strategy, i.e., strategies having
at least one of their properties ($y_k, p_{1k}, p_{2k}, p_{3k}, p_{4k}$) that belongs to the open interval (0, 1) instead of being either 0 or 1:

\[
\vspace{-0.5mm}
\exists k \; | \; \mathcal{S}_{k}
\textnormal{ has at least one of its parameters in } (0,1).
\vspace{-0.5mm}
\]

As pointed out in Section 2, we assume that each player plays according to the memory-one approach.
Then the evolution of the interactions among the strategies adopted by each player can be seen as
a Markov process and, as such, can be completely described by providing
the associate probabilistic transition matrix and the initial state distribution across the state space \cite{Nowak_90, Nowak_90b}.

The matrix of probabilities associated to each possible state (i.e., each
possible outcome of an interaction between two players)
at the $n$-th meeting can be therefore computed 
by multiplying the distribution vector of the 
game initial state  by the transition matrix elevated to $n$-1.

Recalling the formalism of Section 4, we can model an SPDT using the following entities:
\begin{align*}
\mathcal{P} = & \; \{\wp_1, ..., \wp_m\} \hspace{45mm} \textnormal{(the set of players)}\\ 
\mathcal{S}_k = & \; \mathcal{S}(y_k, p_{1k}, p_{2k}, p_{3k}, p_{4k}) \;  \forall k = 1,...,m \hspace{11mm} \left(\begin{matrix} \textnormal{the strategy}\\
\textnormal{of each player}
\end{matrix}\right)\\
\bar{p}_{hk} =  & \; 1-p_{hk} \;\; \forall h = 1,...,4 \;\;\; \forall k = 1,...,m
\hspace{4mm}
\left(\begin{matrix}\textnormal{the complementary}\\ \textnormal{probabilities}
\end{matrix}\right)\\
A_{kj} = & \; \begin{bmatrix} y_k y_j & y_{k}\bar{y}_j & \bar{y}_k y_j & \bar{y}_k\bar{y}_j\end{bmatrix}
\hspace{18mm}
\left(\begin{matrix}
\textnormal{initial distribution}\\
\textnormal{across the states}
\end{matrix}\right)\\
L_{kj} = & \; \begin{bmatrix}
									p_{1k}p_{1j} & p_{1k}\bar{p}_{1j}& \bar{p}_{1k}p_{1j} & \bar{p}_{1k}\bar{p}_{1j} \\
									p_{2k}p_{3j} & p_{2k}\bar{p}_{3j} & \bar{p}_{2k}p_{3j} & \bar{p}_{2k}\bar{p}_{3j} \\
									p_{3k}p_{2j} & p_{3k}\bar{p}_{2j} & \bar{p}_{3k}p_{2j} & \bar{p}_{3k}\bar{p}_{2j} \\
									p_{4k}p_{4j} & p_{4k}\bar{p}_{4j} & \bar{p}_{4k}p_{4j} & \bar{p}_{4k}\bar{p}_{4j}
								\end{bmatrix}
					 \hspace{3.5mm}
					 \left( \begin{matrix}\textnormal{transition matrix}\\ 
					 \textnormal{between } \wp_k \textnormal{ and } \wp_j \end{matrix} \right) \\
Q = & \; \begin{bmatrix} R & S & T & P\end{bmatrix} 
\hspace{38.5mm} \textnormal{(vector of payoffs)}
\end{align*}

\begin{align*}                                
E_{\wp_k}(n) =  & \; \sum_{t=1}^{n} \sum_{\substack{i=1, \\ i \neq k
}}^{m}{A_{ki}L_{ki}^{t-1}Q^{T}}
\hspace{2.5mm} \left( 
\begin{matrix}
\textnormal{expectation of player } \wp_k \textnormal{ after } n\\
 \textnormal{interactions with each other player} \\
\end{matrix}
\right)\\
\end{align*}
We are now ready to compute the
relation $\Delta$, describing the difference between the performances of two generic players $\wp_k$ and $\wp_j$ after $n$ interactions:

\vspace{0.5mm}
\[
\Delta(\wp_k, \wp_j, n) =  \sum_{t=1}^{n} \bigg(\sum_{\substack{i=1, \\ i \neq k}}^{m}{A_{ki}L_{ki}^{t-1}} - \sum_{\substack{i=1,\\ i \neq j}}^{m}{A_{ji}L_{ji}^{t-1}}\bigg) \;Q^{T} =
\]
\begin{equation}
\label{eq:Gamma_Stoc}
= \sum_{t=1}^{n}\Bigg[A_{kj}L_{kj}^{t-1} - A_{jk}L_{jk}^{t-1} + \sum_{\substack{i=1,\\ i\neq j,k}}^{m}(A_{ki}L_{ki}^{t-1} - A_{ji}L_{ji}^{t-1} )\Bigg] \; Q^{T}
\end{equation}
\vspace{0.5mm}

In order to explain what follows, it is worth recalling that every time the matrix $L_{kj}$ described above is
i) irreducible, ii) positive recurrent and iii) aperiodic, then by means of the ergodic theorem it is asymptotically stationary too, i.e.,

\vspace{0.5mm}
\begin{equation}
\exists \;\; \tilde{n} \in \mathbb{N} \;\; \mid \;\; L_{kj}^{\tilde{n}} \; \cong  \; L_{kj}^{\tilde{n}+t} \;\;\;\; \forall k, j = 1,...,m \;, \;\;\; \forall t \in \mathbb{N}
\label{eq:n_accento_circonflesso}
\end{equation}
\vspace{0.5mm}

Then, by means of 
Equation~\eqref{eq:n_accento_circonflesso},
we can approximate 
Equation~\eqref{eq:Gamma_Stoc} as
the sum of two elements, namely
its transitional behavior $F_{kj}$ (i.e. the
sum of the values assumed by the 
function while $t<\tilde{n}$) and its asymptotic behavior $G_{kj}$ (i.e. the sum 
of its value of convergence every
time $t \ge \tilde{n}$). The approximated version of $\Delta$ therefore is:

\vspace{0.5mm}
\begin{equation}
\tilde{\Delta}(\wp_k, \wp_j, n)  \triangleq \left[\;F_{kj}+\sum_{t=\tilde{n}}^{n}G_{kj}\right]Q^{T} =
\left[\;F_{kj}+(n-\tilde{n})G_{kj}\right]Q^{T},
\label{eq:Delta_F_G}
\end{equation}
\vspace{0.5mm}

having defined $F_{kj},\; G_{kj}$ as the next 4-dimensional row vectors:

\vspace{0.5mm}
\[
F_{kj} = \sum_{t=1}^{\tilde{n}-1}\bigg[A_{kj}L_{kj}^{t-1} - A_{jk}L_{jk}^{t-1} + \sum_{\substack{i=1,\\ i\neq j,k}}^{m}\bigg(A_{ki}L_{ki}^{t-1} - A_{ji}L_{ji}^{t-1} \bigg) \bigg]
\]
\vspace{-1mm}
and
\vspace{0.5mm}
\[ 
G_{kj} = \bigg[A_{kj}L_{kj}^{\tilde{n}} - A_{jk}L_{jk}^{\tilde{n}} +  
\sum_{\substack{i=1,\\ i\neq j,k}}^{m} 
\bigg( A_{ki}L_{ki}^{\tilde{n}} - A_{ji}L_{ji}^{\tilde{n}} \bigg)\bigg].\\ 
\]
\vspace{0.5mm}

The approximated version of $\Delta$ can also be expressed in the following alternative way:

\vspace{0.5mm}
\begin{equation}
\tilde{\Delta}(\wp_k, \wp_j, n) = \alpha_{kj} + n\beta_{kj},
\label{eq:Delta_tilde_alpha_beta}
\end{equation}
\vspace{0.5mm}

having defined $\alpha_{kj}=\big[F_{kj} - \tilde{n}G_{kj}\big]Q^T$ and $\beta_{kj} = G_{kj}Q^T$,  $\alpha_{kj}, \beta_{kj} \in \, \mathbb{R}.$ 
We have introduced this additional formulation, because it shows more clearly
that the dependency with $n$ is a straight line where $\beta_{kj}$ is the slope 
and $\alpha_{kj}$ the intercept. Of course, we have multiple straight lines, 
one for each pair of distinct players.
It is worth noting that such lines exist if and 
only if $n > \tilde{n}$ and that $\alpha_{kj}$ and $\beta_{kj}$ 
are linear combinations of the payoffs. 
In the remaining part of the paper we will focus on the case $\beta \neq 0$, which means that 
at each interaction one of the two players does better than the other 
(the case $\beta = 0$ being much less interesting).

\subsection{An Example of Application}
In this subsection we provide an example of a possible real world 
application of the SPDT treated with the GM.

Imagine a scenario where $m$ brands or companies have to 
interact repeatedly in a competitive way. Moreover, let us make some additional
assumptions: i) each interaction can be described as a PD where 
the payoffs are environmental parameters (like exchange fees),
set in some way; ii) the payoffs stay fixed for a given period
of time (for example the calendar year); iii) at the end of this 
period we are interested in the total utility of the entities (earnings of the companies).
From these three assumptions it is clear that we can model the given
scenario as a PD Tournament.

Suppose now that the payoffs can be set only
by a third party, not involved in the interactions. We will 
refer to it as master (as an example, 
the master could be the government or the regulator of the country 
in which the interactions take place). 
Moreover we assume that the master is
somehow related to one of the involved entities and it wants
to set the payoffs \emph{ad-hoc}, in order to guarantee it the highest performance
among all the other ones at the end of the period.  Such specific company might be a state-owned company, to continue with the example.
At the same time the
outperformance gap should not be too wide, in order not to emphasize
the ad-hoc parameters set by the master. Let's say that the gap between the second
best performing entity must not be grater than a threshold $\tau$, again chosen by the master.

Thus the main task of the master is to identify, as a function of
the entities' strategies (which for the sake of simplicity
we suppose known and memory-one), all the possible 4-tuples of values it can use 
to initialize the payoffs in order to achieve its 
goal \emph{with certainty}. Using the model 
provided above and calling $\wp_{*}$ the state-owned company,
the problem can be formalized as follows.

Let us suppose that $\hat{j}$ is the index of the best performing player in $\mathcal{P}$ ($\mathcal{P}$ being the set of all the other companies, i.e., the ones not related to the master):

\vspace{0.5mm}
\[
\hat{j} = \;\; \argmax_{j \in \{1 \ldots m\} } (E_{\wp_j}(n)).
\]
\vspace{0.5mm}

The model $\mathcal{M}$ of the considered problem consists of the following two inequalities:

\vspace{0.5mm}
\begin{equation}
\mathcal{M}(\tau, n) =
\begin{cases}
\Delta(\wp_{*}, \wp_{\hat{j}}, n) > \;0 \\
\Delta(\wp_{*}, \wp_{\hat{j}}, n) < \; \tau
\end{cases}
\label{eq:Model_M_tau_n}
\end{equation}
\vspace{0.5mm}


Once the number of interactions per period has been given
a priori (e.g., $n$ = 100), 
a classical approach is still feasible (we are implicitly assuming that every
entity meets the others the same number of times). 
By arbitrarily imposing $n$ = 100 and $\tau$ = 20, we can solve the 
problem described above, by finding the solutions to the model:

\vspace{0.5mm}
\begin{equation}
\vspace{1mm}
\mathcal{M}(20, 100) =
\begin{cases}
\Delta(\wp_{*}, \wp_{\hat{j}}, 100) > \;0 \\
\Delta(\wp_{*}, \wp_{\hat{j}}, 100) < \; 20
\end{cases}
\label{eq:Gamma_Stoc_100_20}
\vspace{1mm}
\end{equation}
\vspace{0.5mm}

The result are two inequalities with three degrees of freedom, to which
we have to add the three inequalities of the fundamental law, $T>R>P>S$. If the system admits a solution, then it is easy to find it.

However, in real world scenarios, the number of player-player
interactions $n$ is not always known 
in advance.
Unless
some other assumptions on the upper bound of the number 
of interactions are introduced, the only feasible approach is
performing a limit study when $n$ goes to infinity to analyze
$\Delta$'s behavior. Unfortunately this would not allow us to solve
the problem, basically because the gap (if there is) would diverge to either $\pm \infty$.
Thus, within classical GT there is no way to give
an answer that ensures that a specific company will be the best one
with a limiting out-performance threshold.
On the other hand, by using the GM we achieve
a totally different scenario where we are able to \emph{set up} the tournament
and to \emph{precisely analyze} it when $n$ is equal to \G1.
However, when the solution exists, at least one of the payoff has to be chosen
from an infinitely small interval, as stated by the next theorem.

\vspace{0.5mm}
\begin{numbered_theorem}[\textbf{At least one payoff has to be chosen from an infinitely small interval}]

\par
\vspace{5mm}

\par

\textbf{Given}
an SPDT or a DPDT with $m$ players $\wp_1$, $\ldots$, $\wp_m$ and an infinite number of interactions (studied by means of GM)

\textbf{And}
provided a finite upper bound, along with a finite lower bound for the $\tilde{\Delta}_{ij}$ function 
(a function defined over the players $\wp_i$ and $\wp_j$ and over the players' strategies $\mathcal{S}_1$, $\ldots$, $\mathcal{S}_m$)

\textbf{And}
given that at least one relevant payoff is left free to vary (in order to satisfy the constraints on $\tilde{\Delta}_{ij}$)

\textbf{Then\textnormal{,}}
when the set of admissible values for the free payoffs exists, it will contain at least one payoff which has to be chosen from an infinitely small (but still numerically computable) interval.
\end{numbered_theorem}
\vspace{0.5mm}


\begin{proof}
See Appendix B.
\end{proof}

While Appendix B contains a theoretical proof which is valid in general, 
in next section we also provide a constructive proof of the same theorem,
but for a specific numerical example.


\vspace{0.5mm}

\section{Numerical results comparing classical\\
analysis and Grossone Methodology}
We provide now the solution to \eqref{eq:Model_M_tau_n} using both the classical analysis and the GM.
Before doing it, and for the sake of simplicity, let us assume that the number of entities $m$ is equal to 1,
and thus only two strategies are
involved in the competition ($\mathcal{S}_{*}$ and $\mathcal{S}_1$). 
It is worth noting that in this case the payoffs $R$ and $P$ will not play any role in the outcome
of the tournament, since both the players will
obtain $R$ and $P$ on exactly the same occasions.

The numerical results shown
in the following have been computed in Matlab, after
having implemented a simulator for the Infinity 
Computer.

By using the following two strategies:
\[
\vspace{-0.5mm}
\mathcal{S}_{*} = \;\; \mathcal{S}(0.8, 0.75, 0.2, 0.4, 0.05)
\qquad
\mathcal{S}_1 = \;\; \mathcal{S}(0.4, 0.4, 0.1, 0.8, 0.1)
\vspace{0.5mm}
\]
we obtain the initial distribution across the state space:
\[
\vspace{0.5mm}
\resizebox{\linewidth}{!}{
$
A_{*1}  = \left[ \begin{matrix}
0.32 & 0.48 & 0.08 & 0.4 & 0.12 
\end{matrix}
\right]
\qquad
A_{1*}  = \left[ \begin{matrix}
0.32 & 0.08 & 0.48 & 0.8 & 0.12
\end{matrix}
\right]$}
\vspace{0.5mm}
\]
and the following two transition matrices:
\vspace{0.5mm}
\[
\resizebox{\linewidth}{!}{
$L_{*1} = 
\begin{bmatrix}0.3 &	0.45 &	0.1 &	0.15 \\
0.16 &	0.04 &	0.64 &	0.16 \\
0.04 &	0.36 &	0.06 &	0.54 \\
0.005 &	0.045 &	0.095 &	0.855\end{bmatrix}
\qquad
L_{1*} = 
\begin{bmatrix}0.3 &	0.1 &	0.45 &	0.15 \\
0.04 & 0.06 &	0.36 &	0.54 \\
0.16 &	0.64 &	0.04 &	0.16 \\
0.005 &	0.095 &	0.045 &	0.855\end{bmatrix}$}
\vspace{0.5mm}
\]
Their asymptotic counterparts are:
\[
\vspace{0.5mm}
\resizebox{\linewidth}{!}{
$L_{*1}^{\text{asym}} = \begin{bmatrix}
0.038 & 0.106 & 0.148 & 0.708\\
0.038 & 0.106 & 0.148 & 0.708  \\
0.038 & 0.106 & 0.148 & 0.708  \\
0.038 & 0.106 & 0.148 & 0.708 \\\end{bmatrix}
\;\;\;
L_{1*}^{\text{asym}} = \begin{bmatrix}
0.038 & 0.148 & 0.106 & 0.708 \\
0.038 & 0.148 & 0.106 & 0.708 \\
0.038 & 0.148 & 0.106 & 0.708 \\
0.038 & 0.148 & 0.106 & 0.708 \\
\end{bmatrix}$}
\vspace{0.5mm}
\]
We will now study two cases: the case $n=100, \tau = 20$ and the case $n \to \infty, \tau = 20$. The second case will be analyzed using both the limit theory (which as we shall see is not able to find any solution), and the GM (which \emph{is able} to provide a set of solutions, even if the interval from which the parameter $T$ has to be picked from is \emph{infinitely small}).

\subsubsection{Case $n = 100$}
When $n = 100$, the system of inequalities in Equation~\eqref{eq:Model_M_tau_n} becomes:
\[
\vspace{1.5mm}
\mathcal{M}(20, 100) =
\begin{cases}
2.6362 \cdot 10^{29} (T-S) > 0 
\hspace{13.5mm} \textnormal{(always true)} \\
2.6362 \cdot 10^{29} (T-S) < 20 
\hspace{4mm}
\left( \begin{matrix}
   \text{true}\,\text{when}  \\
   T < S+\frac{20}{2.6362 \cdot 10^{29}}  \\
\end{matrix} \right)
\end{cases}
\vspace{1.5mm}
\]
From which we derive the set of all possible 4-tuples solutions:

\vspace{0.5mm}
\[
\hspace{-5mm}
\begin{cases}
T & \in \left(S,\; S+\frac{20}{2.6362 \cdot 10^{29}}\right)\\
R & \in (T, P) \\
P & \in (R, S) \\
S & \text{chosen at will}
\end{cases}
\hspace{+5mm}
\langle \textnormal{\small{\;Solution to}} \;\; \small{\mathcal{M}(20, 100)\;} \rangle
\vspace{1mm}
\]

Observe how the open interval $\left(S,\; S+\frac{20}{2.6362 \cdot 10^{29}}\right)$, from which $T$ has to be picked from, has a finite length in this case.
\vspace{2mm}
\subsubsection{Case $n \to \infty$ \emph{versus} $n=\G1$ (or any other \G1-based infinite number)}
Now we show what we obtain when $n \to \infty$, exploiting respectively the limit theory 
and the GM.

In classical terms, we can see that when $n\to\infty$ the set of the problem's solutions
becomes empty. Indeed, first of all observe that in our toy example, we have
\vspace{0.5mm}
\[
F_{*1} = \left[\begin{matrix}0 & -4.236 \cdot 10^{88} & +4.236 \cdot 10^{88} & 0  \end{matrix}
\right]
\]
\vspace{-0.5mm}
and
\vspace{-0.5mm}
\[
G_{*1} = \left[\begin{matrix}0 & -0.042 & +0.042 & 0 \end{matrix}\right].
\vspace{1.5mm}
\]
Then, applying the limit theory to Equation~\eqref{eq:Delta_F_G} when $\tilde{n}=300$ we get:

\vspace{0.5mm}
\[
\lim_{n\to\infty}\tilde{\Delta}(\wp_{*}, \wp_{\hat{j}}, n) = 
(F_{*1}+ \lim_{n\to\infty} (n-\tilde{n})G_{*1})\;Q^{T} \]
\vspace{0.5mm}

Thus:

\vspace{0.5mm}
\[
\left[ 4.236 \cdot 10^{88} + 0.042 \lim\limits_{n\to\infty}(n-300) \right](T-S) = +\infty.
\]
\vspace{0.5mm}

Note how using $\tilde{n}$ = 300 as asymptotic threshold leads to an an approximation error (with respect to the stationary matrix) close to $10^{-15}$, an acceptable value.

Thus the model $\tilde{\mathcal{M}}$ (which is the approximation of $\mathcal{M}$ when $\tilde{\Delta}$ is used in place of $\Delta$ ) has no solutions, as the second equation is violated by the fact that $\tilde{\Delta}$ tends to the infinity:

\vspace{0.5mm}
\[
\vspace{1mm}
\tilde{\mathcal{M}}(20, n \to \infty) =
\begin{cases}
\lim\limits_{n\to\infty}\tilde{\Delta}(\wp_{*}, \wp_{\hat{j}}, n) > \;0  & \text{(always true)}\\
\lim\limits_{n\to\infty}\tilde{\Delta}(\wp_{*}, \wp_{\hat{j}}, n) < \; 20 & \text{(always false)}
\end{cases}
\]
\vspace{0.5mm}

The fact that $\tilde{\Delta}(\wp_{*}, \wp_{\hat{j}}, n)$ tends to infinity is not accidental, since its asymptotic behavior does not depend on the value
of $\tilde{n}$, nor on the values of $T$ and $S$.

This is even more apparent when using for $\tilde{\Delta}$ the expression given on Equation~\eqref{eq:Delta_tilde_alpha_beta}:
\[
\vspace{0mm}
\tilde{\Delta}(\wp_*, \wp_{\hat{j}}, n) = \alpha_{*\hat{j}} + n \beta_{*\hat{j}}
\vspace{+1.0mm}
\]
from which the fact that $\tilde{\Delta}$ grows linearly with $n$ is obvious, and thus diverges (the only exception is when $\beta_{*\hat{j}} = 0$, a rare event).

In conclusion there is no way to set the four
parameters ($T$, $R$, $P$ and $S$) in order to satisfy the second inequality in Equation~\eqref{eq:Model_M_tau_n}. \\

On the other hand, using the \emph{Grossone} approach, the divergence is avoided by the fact that \G1 is a concrete number and, therefore, numerical computations can be executed with it. Thus if the solutions exist they can still be numerically computed:
\vspace{0.5mm}
\begin{align*}
\Delta(\wp_{*}, \wp_1, \G1) = & \;\; (F_{*1}+ (\G1-\tilde{n})G_{*1})\;Q^{T} \\
= & \;\; 4.2362 \cdot 10^{88} + 0.0416(\G1-300)](T-S).
\end{align*}
\vspace{-0.5mm}
Considering the given numerical example, the solving system\\ $\tilde{\mathcal{M}}(20, \G1)$ becomes:
\vspace{0.5mm}
\[
\begin{cases}
\tilde{\Delta}(\wp_{*}, \wp_{\hat{j}}, \G1) > 0 \hspace{38mm} \textnormal{(always true)}\\
\tilde{\Delta}(\wp_{*}, \wp_{\hat{j}}, \G1) < 20 \hspace{18mm} \left( \begin{matrix}
   \text{true}\,\text{when}  \\
   T < S + \frac{20}{4.2362 \cdot 10^{88} + 0.0416\left(\G1-300\right)}  \\
\end{matrix} \right)
\end{cases}
\]
\vspace{3.5mm}
The associated set of all possible 4-tuples solving it is:
\vspace{-1.5mm}
\begin{equation}
 \hspace{-.5mm}
\begin{cases}
T & \in \left( S,\; S + \frac{20}{4.2362 \cdot 10^{88} + 0.0416\left(\G1-300\right)} \right)\\
R & \in (T, P) \\
P & \in (R, S) \\
S & chosen\; at\; will
\end{cases}
\hspace{-3mm}\langle \textnormal{\small{\;Solution to}} \;\; \footnotesize{\tilde{\mathcal{M}}(20, \G1)\;}  \rangle
\label{eq:solutions_to_M_20_grossone}
\end{equation}

\vspace{2mm}

The direct consequence of the previous considerations is that, exploiting
the GM, we can \emph{always} guarantee the possibility of creating an environment
(opportunely tuning the free parameters) where the victory of one strategy
is assured independently from the specific value assigned to the
number of interactions and to the maximum
size of the margin of victory. Continuing the analogy with 
the example, we are able to find the whole set of 4-tuples ($T$, $R$, $P$, $S$)
such that our goal is assured to be reached independently from how
large or small $n$ and $\tau$, respectively, could
be.

Observe how this result is not contradictory with classical limit theory. Indeed, the open interval $\left( S,\; S + \frac{20}{4.2362 \cdot 10^{88} + 0.0416\left(\small{\G1}-300\right)} \right)$, from which $T$ has to be picked from, is \emph{infinitely small}. Using the classical limit theory we have not been able to derive (and even describe) the set of solutions provided in Equation~\eqref{eq:solutions_to_M_20_grossone}, because of the intrinsic limitation of the mathematical language used therein.

Finally we wish to point out that the same numerical asymptotic analysis based on IC could be performed even using different \G1-based numbers of interactions, such as $n=\frac{\small{\G1}}{2}$, $n=\frac{\small{\G1}}{3}$, etc., especially when this choice is appropriate to model a peculiar feature of the SPDT at hand. This is another kind of study that cannot be performed using classical limit theory.

\section{Conclusions}
In this paper we have shown how, by using GM,
it is possible to numerically build and solve new
classes of PD Tournaments which were even difficult to imagine and/or formulate,
up to date.
Moreover, not only the GM allowed us to formulate such problems
but also to solve them \emph{numerically} in Matlab, i.e., made this kind of 
problems \emph{computationally} 
solvable even if they contain 
finite, infinite and infinitesimal 
quantities at the same time.






\begin{appendices}
\input{appendix_A.tex}

\input{appendix_B.tex} 
\end{appendices}

\section*{Acknowledgements}
The authors thank the three unknown reviewers for their valuable comments.

\vspace{+5mm}
\bibliographystyle{ijuc} 
\bibliography{bibGrossGameTheory}

\end{document}

%% file: appendix_A.tex
\section{Expectations Given on Table 1}


In this appendix we provide a complete step-by-step explanation
of the equations provided in Table 1. 
The explicit and simple form in which the expectation of each player can be
written is due to the fact that we restrict ourselves to a deterministic
tournament and therefore to a deterministic evolution of the game's
history.

Let us start analyzing  the player by player interactions and
then to sum up the results obtained in the total expectation
of each strategy.

\subsection{Interactions between $\wp_1$ and $\wp_2$}

At the very first interaction between the two players, $\wp_1$ defects
and $\wp_2$ cooperates. Therefore $\wp_1$ earns $T$ and $\wp_2$ $S$. Moreover,
because of $\wp_1$ defection, $\wp_2$ will defect at every other interaction
with $\wp_1$.

At the second interaction and for all the subsequent ones both $\wp_1$ and
$\wp_2$ defect, ending both to gain $P$ each time. More precisely, calling
$n$ the total number of interactions between $\wp_1$ and $\wp_2$, this event
(the mutual defection) happens $n$-1 times, since the first interaction differs from the others.

Summing up, the expectations of the two players
after $n$ interactions are:
\[
\vspace{-4mm}
E_{\wp_1,\wp_2} (n) = T + (n-1)P \qquad E_{\wp_2,\wp_1}(n) = S + (n-1)P
\vspace{5mm}
\]

\subsection{Interactions between $\wp_1$ and $\wp_3$}
At the very first interaction between the two players, $\wp_1$ defects
and $\wp_3$ cooperates. Therefore $\wp_1$ earns $T$ and $\wp_3$ $S$. 
Moreover,
because of $\wp_1$'s defection and remembering that $\wp_2$ subsequent 
choice is its enemy's previous one, $\wp_2$ next action against $\wp_1$ 
will be a defection.

At the second interaction both $\wp_1$ and $\wp_3$ defect, ending both 
to gain $P$. It is worth noticing that this event (the mutual
defection) will be the outcome of every other interaction between 
$\wp_1$ and $\wp_3$, being $\wp_3$'s behavior dependent on $\wp_1$'s one 
which is fixed. More precisely, calling
$n$ the total number of interactions between $\wp_1$ and $\wp_3$, the 
mutual defection happens $n$-1 times,
since the first interaction differs from all subsequent ones, which are identical.

Summing up what just said, the expectations of the two players
after their $n$ interactions are:
\[
\vspace{-4mm}
E_{\wp_1, \wp_3}(n) = T + (n-1)P \qquad E_{\wp_3,\wp_1}(n) = S + (n-1)P
\vspace{5mm}
\]
\subsection{Interactions between $\wp_2$ and $\wp_3$}

At the very first interaction between the two players, both $\wp_2$ 
and $\wp_3$ cooperate. Therefore both earn $R$. Moreover, because 
$\wp_3$ copies $\wp_2$'s previous behavior and $\wp_2$ continues to 
cooperate as long as $\wp_3$ does the same, it is quite obvious that 
all the subsequent interactions between such two players will result 
in a mutual cooperation (i.e., both players gain $R$).

In conclusion, at every interaction between $\wp_2$ and $\wp_3$ the 
outcome is a mutual cooperation. This leads an expectation for both players equal to:
\[
\vspace{-4mm}
E_{\wp_2,\wp_3}(n) = E_{\wp_3,\wp_2}(n) = n R
\vspace{5mm}
\]
where again $n$ is the total number of interactions between the two players.

\subsection{Total expectations}

In this sub-section we briefly show the resulting total expectation
of the three players after exactly $n$ interactions with each other.
These results can be easily obtained by summing up the preceding expectations, player by player:
\vspace{0.5mm}
\begin{table}[ht]
\begin{center}
\begin{tabular}{r l} 
  $E_{\wp_1}(n)$  =  &  $E_{\wp_1,\wp_2}(n) + E_{\wp_1,\wp_3}(n) \; = \; 2\left(T + (n-1)P\right)$ \\
  $E_{\wp_2}(n)$  =  &  $E_{\wp_2,\wp_1}(n) + E_{\wp_2,\wp_3}(n) \; = \; nR + (n-1)P + S$ \\
  $E_{\wp_3}(n)$  =  &  $E_{\wp_3,\wp_1}(n) + E_{\wp_3,\wp_2}(n) \; = \; nR + (n-1)P + S$
\end{tabular}
\end{center}
\end{table}
\vspace{-2mm}
\FloatBarrier

%% file: appendix_B.tex
\section{Proof of Theorem 1}

To prove Theorem 1, let us first define the following \emph{positive shift function} $\Psi_\delta(x)$, 
which adds a positive value $\delta$ to x:
\[
  \Psi_{\delta}(x) = x + \delta,  \hspace{5mm} \delta > 0  
\]
The fundamental law system
\vspace{0.5mm}
\[
F(T,R,P,S) = 
\begin{cases}
T > R \\
R > P \\
P > S
\end{cases}
\]
\vspace{0.5mm}
can now be rewritten using $\Psi$ as:
\vspace{0.5mm}
\begin{equation}
F(T,R,P,S)= 
\begin{cases}
T = \Psi_{\delta_T'}(R) = R + \delta_T' , & ( \delta_T' > 0, \; \textnormal{free to vary} )\\
R = \Psi_{\delta_R'}(P) = P + \delta_R' , & ( \delta_R' > 0, \; \textnormal{free to vary} )\\
P = \Psi_{\delta_P'}(S) = S + \delta_P' , & ( \delta_P' > 0, \; \textnormal{free to vary} )
\end{cases}
\label{eq:eq_Fundamental_Law_param}
\end{equation}
\vspace{0.5mm}
From \eqref{eq:eq_Fundamental_Law_param} we get that three payoffs over four can be always described as a function of the fourth,
using the positive shift functions:
\vspace{0.5mm}
\begin{equation}
F(S,\delta_P,\delta_R,\delta_T) = 
\begin{cases}
T = \Psi_{\delta_T'}(\Psi_{\delta_R'}(\Psi_{\delta_P'}(S))) = \Psi_{\delta_T}(S) = S + \delta_T\\
R = \Psi_{\delta_R'}(\Psi_{\delta_P'}(S)) = \Psi_{\delta_R}(S) = S +\delta_R\\
P = \Psi_{\delta_P'}(S) = \Psi_{\delta_P}(S) = S + \delta_P
\end{cases}
\label{eq:eq_Fundamental_Law_paramS}
\end{equation}
\vspace{0.5mm}

where $\delta_T$, $\delta_R$ and $\delta_P$ are defined as functions of $\delta_T'$, $\delta_R'$ and $\delta_P'$ defined above:
\vspace{0.5mm}
\[
\delta_T = \delta_T' + \delta_R' + \delta_P
\]
\[
\delta_R = \delta_R' + \delta_P
\]
\[
\delta_P = \delta_P'
\]
\vspace{0.5mm}

Combining \eqref{eq:eq_Fundamental_Law_paramS} and \eqref{eq:Delta_tilde_alpha_beta} we get
the following alternative expressions for $\alpha_{kj}$ and $\beta_{kj}$, as functions of $S$, $\delta_P$, $\delta_R$, and $\delta_T$:

\vspace{0.5mm}
\[
   \alpha_{kj} = \lambda_{kj}^1 \Psi_{\delta_R}(S) + \lambda_{kj}^2 S + \lambda_{kj}^3 \Psi_{\delta_T}(S) + \lambda_{kj}^4 \Psi_{\delta_P}(S)
\]
\[
\beta_{kj} = \mu_{kj}^1 \Psi_{\delta_R}(S) + \mu_{kj}^2 S + \mu_{kj}^3 \Psi_{\delta_T}(S) + \mu_{kj}^4 \Psi_{\delta_P}(S)
\]
\vspace{0.5mm}

where $\lambda_{kj}^i$ and $\mu_{kj}^i$ are the i-th component of the vector ($F_{kj}-\tilde{n}G_{kj}$) and $G_{kj}$, respectively.

Thus now \eqref{eq:Delta_tilde_alpha_beta} can be rewritten explicitly with respect to $S$ as:

\vspace{0.5mm}
\begin{equation}
    \tilde{\Delta}(\wp_k, \wp_j, n) = 
    (\gamma_1 n + \gamma_2) S + \gamma_3 n + \gamma_4
\label{eq:Delta_paramS}
\end{equation}
\vspace{0.1mm}
where
\vspace{0.1mm}
\[\gamma_1 = \sum_{i = 1}^4 \mu_{kj}^i, \qquad \gamma_3 = \mu_{kj}^1 \delta_R + \mu_{kj}^3 \delta_T + \mu_{kj}^4 \delta_P\]
\[\gamma_2 = \sum_{i = 1}^4 \lambda_{kj}^i, \qquad \gamma_4 = \lambda_{kj}^1 \delta_R + \lambda_{kj}^3 \delta_T + \lambda_{kj}^4 \delta_P\]
\vspace{0.5mm}

Hereinafter we have to distinguish two cases: $\gamma_1 \neq 0$ and $\gamma_1 = 0$. Let us start from the first one.\\

\subsection{Case $\gamma_1 \neq 0$}

By means of \eqref{eq:Delta_paramS} we can write the approximated version of \eqref{eq:Model_M_tau_n} when $n$ = \G1 and $\tau$ is finite, in the following two alternative ways, according to the sign of $\gamma_1$:

\vspace{0.5mm}
\begin{equation}
\tilde{\mathcal{M}}(\G1, \tau) =
\begin{cases}
     S > - \frac{\gamma_3 \footnotesize{\G1} + \gamma_4}{\gamma_1 \footnotesize{\G1} + \gamma_2} \\
     S < \frac{\tau - \gamma_3 \footnotesize{\G1} - \gamma_4}{\gamma_1 \footnotesize{\G1} + \gamma_2}
\end{cases}
\textnormal{iff} \;\;\, \gamma_1 > 0
\label{eq:Model_M_tau_n_S_pos}
\end{equation}
\vspace{0.5mm}
or
\vspace{0.5mm}
\begin{equation}
\tilde{\mathcal{M}}(\G1, \tau) =
\begin{cases}
     S < - \frac{\gamma_3 \footnotesize{\G1} + \gamma_4}{\gamma_1 \footnotesize{\G1} + \gamma_2} \\
     S > \frac{\tau - \gamma_3 \footnotesize{\G1} - \gamma_4}{\gamma_1 \footnotesize{\G1} + \gamma_2}
\end{cases}
\textnormal{iff} \;\;\, \gamma_1 < 0
\label{eq:Model_M_tau_n_S_neg}
\end{equation}
\vspace{0.5mm}


Looking at \eqref{eq:Model_M_tau_n_S_pos} we can deduce that at least the payoff $S$ must be chosen in an infinitely small interval. 
Indeed

\vspace{0.5mm}
\[S \in \left(- \frac{\gamma_3 \footnotesize{\G1} + \gamma_4}{\gamma_1 \footnotesize{\G1} + \gamma_2},  \frac{\tau - \gamma_3 \footnotesize{\G1} - \gamma_4}{\gamma_1 \footnotesize{\G1} + \gamma_2}\right),\]
\vspace{0.5mm}

an interval whose width $W$ is equal to:

\vspace{0.5mm}
\begin{equation}
W = \frac{\tau - \gamma_3 \footnotesize \footnotesize{\G1} - \gamma_4}{\gamma_1 \footnotesize{\G1} + \gamma_2} + \frac{\gamma_3 \footnotesize{\G1} + \gamma_4}{\gamma_1 \footnotesize{\G1} + \gamma_2} = \frac{\tau}{\gamma_1 \footnotesize{\G1} + \gamma_2}
\label{eq:interval_W1}
\end{equation}
\vspace{0.5mm}
Such width is infinitesimal, since both $\gamma_1$ and $\gamma_2$ are finite.
The same considerations hold for \eqref{eq:Model_M_tau_n_S_neg}, when using for $S$ the interval:

\vspace{0.5mm}
\[
S \in 
\left(\frac{\tau - \gamma_3 \footnotesize{\G1} - \gamma_4}{\gamma_1 \footnotesize{\G1} + \gamma_2}, - \frac{\gamma_3 \footnotesize{\G1} + \gamma_4}{\gamma_1 \footnotesize{\G1} + \gamma_2}\right).
\]
\vspace{0.5mm}

\subsection{Case $\gamma_1 = 0$}

The fact that $\gamma_1 = 0$ has two consequences: 
\begin{enumerate}
    \item either $S$ can be chosen in a finite interval (this happens every time $\gamma_2 \neq 0$, see \eqref{eq:interval_W1}) or it is free to vary, being problem-independent (this happens when $\gamma_2 = 0$).
    Thus in the following we can assume that $S$ has been fixed, by choosing it at will or from within the finite length interval;
    \item at least one among $\mu_{kj}^1, \mu_{kj}^3, \mu_{kj}^4$ is non-zero.
\end{enumerate}

The second assertion is true because if they were all equal to zero, $\mu_{kj}^2$ would be zero too (since $\gamma_1 = 0$). Then, by definition, $\beta_{kj}$ would be zero, a case we decided not to analyze, being it uninteresting. Thus we may rewrite \eqref{eq:Delta_paramS} explicitly with respect to a $\delta$ associated to a non-zero $\mu_{kj}$, let it be $\delta_T$. The idea is to prove that in this case it is the payoff $T$ which
has to vary within an infinitely small interval. Indeed the approximated $\Delta$ function can be now rewritten as:

\vspace{0.5mm}
\begin{equation}
    \tilde{\Delta}(\wp_k, \wp_j, n) = \delta_T(\mu_{kj}^3 n + \lambda_{kj}^3) + \gamma_2 S + \gamma_5 n + \gamma_6
\label{eq:Delta_param_deltaT}
\end{equation}
\vspace{0.5mm}

where
\vspace{0.5mm}
\[\gamma_5 \; = \; \gamma_3 - \mu_{kj}^3 \delta_T \; = \;  \mu_{kj}^1 \delta_R + \mu_{kj}^4 \delta_P\]
\[\gamma_6  \; = \; \gamma_4 - \lambda_{kj}^3 \delta_T \; = \; \lambda_{kj}^1 \delta_R + \lambda_{kj}^4 \delta_P\]
\vspace{0.5mm}

By means of \eqref{eq:Delta_param_deltaT} we can write the approximated version of \eqref{eq:Model_M_tau_n} when $n$ = \G1 and $\tau$ is finite, as follows:

\vspace{0.5mm}
\begin{equation}
\tilde{\mathcal{M}}(\G1, \tau) =
\begin{cases}
     \delta_T > - \frac{\gamma_5 \footnotesize{\G1} + \gamma_6 + \gamma_2 S}{\mu_{kj}^3  \footnotesize{\G1} + \lambda_{kj}^3} \\
     \delta_T < \frac{\tau - \gamma_5 \footnotesize{\G1} - \gamma_6 - \gamma_2 S}{\mu_{kj}^3  \footnotesize{\G1} + \lambda_{kj}^3}
\end{cases}
\textnormal{iff} \;\;\, \mu_{kj}^3 > 0
\label{eq:Model_M_tau_n_deltaT_pos}
\end{equation}
\vspace{0.5mm}

or

\vspace{0.5mm}
\begin{equation}
\tilde{\mathcal{M}}(\G1, \tau) =
\begin{cases}
     \delta_T < - \frac{\gamma_5 \footnotesize{\G1} + \gamma_6 + \gamma_2 S}{\mu_{kj}^3  \footnotesize{\G1} + \lambda_{kj}^3} \\
     \delta_T >\frac{\tau - \gamma_5 \footnotesize{\G1} - \gamma_6 - \gamma_2 S}{\mu_{kj}^3  \footnotesize{\G1} + \lambda_{kj}^3}
\end{cases}
\textnormal{iff} \;\;\, \mu_{kj}^3 < 0
\label{eq:Model_M_tau_n_deltaT_neg}
\end{equation}
\vspace{0.5mm}

Looking at \eqref{eq:Model_M_tau_n_deltaT_pos} we can deduce that $\delta_T$, once fixed $S$, 
must be chosen from within an infinitely small interval.
Indeed 

\vspace{0.5mm}
\begin{equation}
\delta_T \in \left(- \frac{\gamma_5 \footnotesize{\G1} + \gamma_6 + \gamma_2 S}{\mu_{kj}^3  \footnotesize{\G1} + \lambda_{kj}^3}, \;  \frac{\tau - \gamma_5 \footnotesize{\G1} - \gamma_6 - \gamma_2 S}{\mu_{kj}^3  \footnotesize{\G1} + \lambda_{kj}^3}\right),
\label{eq:deltaT_interval}
\end{equation}
\vspace{0.5mm}

i.e., an interval having width $W$ equal to:

\vspace{0.5mm}
\begin{equation}
W = \frac{\tau - \gamma_5 \footnotesize{\G1} - \gamma_6 - \gamma_2 S}{\mu_{kj}^3  \footnotesize{\G1} + \lambda_{kj}^3} + \frac{\gamma_5 \footnotesize{\G1} + \gamma_6 + \gamma_2 S}{\mu_{kj}^3  \footnotesize{\G1} + \lambda_{kj}^3} = \frac{\tau}{\mu_{kj}^3  \footnotesize{\G1} + \lambda_{kj}^3}
\label{eq:interval_W2}
\end{equation}
\vspace{0.5mm}

Such width is infinitesimal, because both $\mu_{kj}^3$ and $\lambda_{kj}^3$ are finite. Moreover, combining the first 
equation of system \eqref{eq:eq_Fundamental_Law_paramS} with \eqref{eq:deltaT_interval} we get the interval from within $T$ has to vary:

\vspace{0.5mm}
\[T \in \left(S - \frac{\gamma_5 \footnotesize{\G1} + \gamma_6 + \gamma_2 S}{\mu_{kj}^3  \footnotesize{\G1} + \lambda_{kj}^3}, \; S + \frac{\tau - \gamma_5 \footnotesize{\G1} - \gamma_6 - \gamma_2 S}{\mu_{kj}^3  \footnotesize{\G1} + \lambda_{kj}^3}\right)\]
\vspace{0.5mm}

that is again an infinitesimal interval with the same width given in \eqref{eq:interval_W2}.

The same considerations hold for \eqref{eq:Model_M_tau_n_deltaT_neg}, but using the next interval for $\delta_T$:

\vspace{0.5mm}
\[
\delta_T \in \left(\frac{\tau - \gamma_5 \footnotesize{\G1} - \gamma_6 - \gamma_2 S}{\mu_{kj}^3  \footnotesize{\G1} + \lambda_{kj}^3}, - \frac{\gamma_5 \footnotesize{\G1} + \gamma_6 + \gamma_2 S}{\mu_{kj}^3  \footnotesize{\G1} + \lambda_{kj}^3}\right).   \;\;\; \boxempty
\]
\vspace{0.5mm}

It is worth noting that the width of the infinitely small interval does not depend on the effective and subsequent choice of the other payoffs', i.e., it does not depend on the values assigned to either $\delta_T, \delta_R, \delta_P$ (when $\gamma_1 \neq 0$, as it can be seen from \eqref{eq:interval_W1}) or $S, \delta_R, \delta_P$ (when $\gamma_1 = 0$, as it can be seen from \eqref{eq:interval_W2}).